\documentclass{appolb}
\usepackage{epsfig}

\usepackage[T1]{fontenc}
\usepackage[cp1250]{inputenc}
\usepackage{graphicx}
\usepackage{psfrag}
\usepackage{bbm}
\usepackage{amsmath}

\newcommand{\bq}{\begin{equation}}
\newcommand{\eq}{\end{equation}}
\newcommand{\ba}{\begin{eqnarray}}
\newcommand{\ea}{\end{eqnarray}}

\begin{document} 

\title{Random matrices and localization in the quasispecies theory
\thanks{Presented at the 23rd Marian Smoluchowski Symposium on Statistical Physics -- Random Matrices, Statistical Physics and Information Theory, 26-30 September 2010, Kraków, Poland.}
}

\author{B. Waclaw
\address{School of Physics and Astronomy, University of Edinburgh, Mayfield Road, Edinburgh EH9 3JZ, United Kingdom}
}
\maketitle
\begin{abstract}
The quasispecies model of biological evolution for asexual organisms such as bacteria and viruses has attracted considerable attention of biological physicists. Many variants of the model have been proposed and subsequently solved using the methods of statistical physics. In this paper I will put forward important but largely overlooked relations between localization theory, random matrices, and the quasispecies model. These relations will help me to study the dynamics of this model. In particular, I will show that the distribution of times between evolutionary jumps in the genotype space follows a power law, in agreement with recent findings in the shell model -- a simplified version of the quasispecies model.
\end{abstract}
\PACS{87.23.Kg, 87.10.-e, 02.50.-r, 05.70.Fh}

\section{Introduction}
The theory of biological evolution is a pillar of modern biology and has been inspiring scientists since the time of Darwin. Evolution operates by two primary processes: natural selection, in which best adapted organisms outcompete less adapted ones, and variability in different traits of individual organisms, which are passed down from generation to generation. Although natural selection was supported by rather strong experimental evidence already 150 years ago, it was only in 1950's when the discovery of the role of DNA in heredity and, subsequently, the explanation of its molecular structure provided a molecular basis for genetic variability. Today we know that this variability is related to changes to the genetic material stored in DNA, either by DNA exchange in sexual reproduction or microbial conjugation, or by mutations -- random changes in the sequence of nucleotides which form the DNA chain.

All organisms capable of self-reproduction which exist today are complicated systems of many coupled chemical reactions, usually taking place in isolated compartments -- cells -- or even smaller regions within cells. Is evolution a phenomenon which started to operate after living organisms had emerged a few billion of years ago, or was it preceded by a similar process acting on molecules floating freely in the oceans? In an attempt to answer that question, Eigen and Schuster conceived a theoretical model \cite{eigen} explaining how selection and mutation could work already on the level of single chemical molecules. In their model, macromolecules such as DNA or RNA, which are linear polymers composed of nucleotides, were subjected to error-prone replication. If errors (mutations) were rare, the molecule with the highest replication rate soon dominated the population. For increasing mutation rate, however, the fittest molecule was surrounded by an expanding cloud of mutants in the space of all possible sequences. Eigen and Schuster called this cloud ``quasispecies'', by analogy to the concept of a biological species, which consists of closely related genotypes. 

The model predicts that if the mutation rate increases beyond some critical value -- the error threshold -- the quasispecies becomes ``delocalized''. This means that the fittest molecule corresponding to some particular sequence of nucleotides is lost and all possible sequences start to appear. If these other sequences are much less fit or even incapable of reproduction (lethal mutations), the population will inevitably die out. This is called the error catastrophe. The error threshold is predicted to decrease with increasing length of the sequence, suggesting that for a given mutation rate, the amount of genetic information stored in a self-replicating molecule is restricted. On the other hand, higher mutation rate means improved adaptability to changing conditions. Indeed, it has been found experimentally \cite{nowak}, that the evolution of some viruses such as HIV operates very close to the 
error threshold.

The molecular quasispecies theory has attracted considerable attention not only from biologists but also mathematicians and physicists. In particular, the quasispecies model has been studied by mapping it onto Ising spin chains \cite{leithausser,demetrius,baake,baake2}, directed polymers \cite{kussel}, or, more recently, Anderson localization \cite{epstein,BW}, and has been solved exactly in some special cases \cite{baake,gallucio, woodcock}. The purpose of this work is to re-examine the quasispecies model, or, more precisely, its version with parallel mutation and selection \cite{baake}, and to show an analogy between this model and random matrices which appear in the theory of Anderson localization. I will first define the quasispecies model and discuss its several versions which can be found in the literature. Next, I will show how some questions fundamental to the quasispecies theory can be rephrased using the language of random matrix theory (RMT). Finally, I will show how to answer these questions using some simple concepts borrowed from RMT. 

\section{Quasispecies model}
The quasispecies model \cite{eigen} describes biological evolution of simple organisms which reproduce asexually in a chemostat -- a bioreactor with constant supply of fresh nutrients and removal of liquid culture and waste products, so that the culture volume is kept constant. Every organism has a ``genotype'' which is a sequence of length $L$ composed of symbols taken from some finite alphabet. The symbols could correspond to four different nucleotides which are the building blocks of DNA, but one usually considers binary sequences composed of only two symbols $0$ and $1$. This assumption simplifies calculations but it has no qualitative effect on any properties of the model I will mention in this manuscript. The number of all possible genotypes is $N=2^L$. Each genotype can be labelled by an integer number $i=1,\dots,N$, and the binary representation of $i-1$ gives the corresponding binary sequence. 

The genotypes reproduce by replication. However, the copy procedure is not error-free: each symbol has a finite probability $\gamma$ of being substituted by another (randomly chosen) symbol in the process of replication. Assuming that errors are made independently at any position in the sequence, the matrix $Q_{ji}$ which describes the probability of mutation from genotype $i$ to genotype $j$ reads
\bq
	Q_{ji} = \gamma^{d(i,j)} (1-\gamma)^{L-d(i,j)}, \label{qdef}
\eq
where $d(i,j)$ is the Hamming distance between the genotypes $i,j$, i.e., $d(i,j)$ is the number of positions at which the corresponding sequences are different. For $j=i$, $Q_{ii}=(1-\gamma)^L$ gives the probability of replicating without errors. By definition, $Q_{ji}$ is a symmetric, square, doubly stochastic matrix: $\sum_j Q_{ji}=\sum_i Q_{ij} = 1$. 

If we denote by $\phi_i$ the specific growth rate of genotype $i$, we can describe the model by the following reactions 
\bq
	\mbox{genotype}\; i \xrightarrow{\phi_i} \left\{
\begin{array}{l}
	\mbox{genotype}\; i + \mbox{genotype}\; i \qquad (\mbox{with probability}\; Q_{ii}) \\
	\mbox{genotype}\; i + \mbox{genotype}\; j \qquad (\mbox{with probability}\; Q_{ji}) 
\end{array} \right. \label{quasireact}
\eq
in which $\phi_i$ above the arrow means that the reaction constant is $\phi_i$. Suppose now that the population is very large so that we can neglect stochastic fluctuations of the numbers of genotypes. Then, the time evolution of the abundances (number densities) $n_1,\dots,n_N$ of genotypes $i=1,\dots,N$ is modelled by the set of $N$ differential equations:
\bq
	\frac{{\rm d}}{{\rm d}t} n_i(t) =  \sum_{j=1}^N Q_{ij}\phi_j n_j(t) - n_i(t) J(t), \label{quasieq1}
\eq
where $J(t)$ is the rate at which organisms (molecules) are washed out from the system. The term $J(t)$ forces the system to evolve towards the steady state, otherwise the growth would always be exponential. In this paper, $J(t)$ is assumed to be proportional to the overall concentration $\sum_i n_i(t)$. This causes the net growth rates to become negative if the population is too dense, as if the organisms competed for limited resources. 

Since $J(t)$ depends on $n_i(t)$, the quasispecies equation (\ref{quasieq1}) is non-linear. However, a simple change of variables:
\bq
	x_i(t)=n_i(t)\exp\left(\int_0^t J(t') dt'\right) \label{ntox}
\eq
reduces Eq.~(\ref{quasieq1}) to a linear equation,
\bq
	\frac{{\rm d}}{{\rm d}t} x_i(t) = \sum_{j=1}^N Q_{ij}\phi_j x_j(t).
\eq
As we are usually interested only in the ratios of different $n_i$'s (relative concentrations of genotypes), we can use $x_i$ instead of $n_i$ to describe the state of the system. The above equation can be rewritten as
\bq
	\frac{{\rm d}}{{\rm d}t}\vec{x}(t) = W \vec{x}(t), \label{model0}
\eq
where the matrix $W_{ij}=Q_{ij} \phi_j$. Let $\lambda_1>\lambda_2>\dots>\lambda_N$ be the set of eigenvalues of $W$, and $\{\vec{\psi_i}\}$ denote the corresponding eigenvectors:
\bq
	W\vec{\psi_i} = \lambda_i \vec{\psi_i}.
\eq
Then, Eq.~(\ref{model0}) has the following solution
\bq
	\vec{x}(t) = e^{tW}\vec{x}(0)= \sum_i e^{t\lambda_i} \left(\vec{\psi_i}\cdot\vec{x}(0)\right) \vec{\psi}_i, \label{quasiformal}
\eq
which for large times reduces to $\vec{x}(t\to\infty) \propto \vec{\psi}_1$. Therefore, the steady state of Eq.~(\ref{quasieq1}) is proportional to the eigenvector $\vec{\psi_1}$ to the largest eigenvalue $\lambda_1$ of the matrix $W$. This gives the steady-state abundances
\bq
	\vec{n}^* \equiv \vec{n}(t\to\infty) = \frac{\lambda_1}{\sum_i \psi_{1,i}} \vec{\psi}_1. \label{vecnn}
\eq
The physical properties of the formal solution from Eq.~(\ref{quasiformal}) and Eq.~(\ref{vecnn}) depend on the choice of the growth rates $\{\phi_i\}$, which specify ``fitnesses'' of different genotypes, i.e., how well they are adapted to the environment. 
The graph of possible mutations plus the fitnesses is usually referred to as the fitness landscape. This metaphor is based on viewing fitness peaks as mountains of different heights, with the population climbing generally uphill in the course of evolution and moving from lower to higher peaks, until the highest peak (global fitness maximum) is reached. Although frequently used in population biology, the concept of fitness has had an important drawback: until recently little information was available on real fitness landscapes because experimental evaluation of the fitness for large numbers of genotypes is very difficult. Lacking experimental data, many models for the fitness landscape have been considered, without a priori knowledge of which one is correct. Some of the most popular choices are listed below:
\begin{enumerate}
	\item single-peak landscape \cite{gallucio}: the fitness $\phi_1$ of a single genotype is taken to be maximal, and $\phi_2=\dots=\phi_{N}$ are assumed to be smaller than $\phi_1$. This corresponds to a situation in which there is one best-fit genotype (master sequence) and all mutants are less fit.
	\item multiplicative landscape \cite{woodcock}: $\phi_1$ is maximal, and $\phi_i=\phi_1 (1-s)^{d(1,i)}$ where $s$ is some positive constant and $d(1,i)$ is the Hamming distance between the sequences $1$ and $i$. In this model, each single-symbol mutation lowers the fitness.
	\item ``holey'' landscape: the fitness of each genotype is either large (fit genotypes) or small (unfit genotypes). Thus, in the fitness landscape, there are holes of unfit genotypes surrounding islands of equally fit genotypes. The fitnesses can be either correlated \cite{gavrilets} or uncorrelated \cite{gavrilets2}.
	\item rugged fitness landscape \cite{kauffman,jain}: the fitnesses are drawn as independent, identically distributed random numbers from some continuous distribution $p(\phi)$, the same for all genotypes. 
\end{enumerate}
Most analytical results has been obtained for the single-peaked landscape (1), whereas the most realistic one is probably the rugged landscape (4). However, what all these landscapes have in common is the emergence of localized ``quasispecies'' and (for some of them) the existence of the error threshold.

\begin{figure}
\psfrag{N_i}{$n_i^*$}\psfrag{i}{$i$}\psfrag{g}{$\gamma$}
\includegraphics[width=5cm]{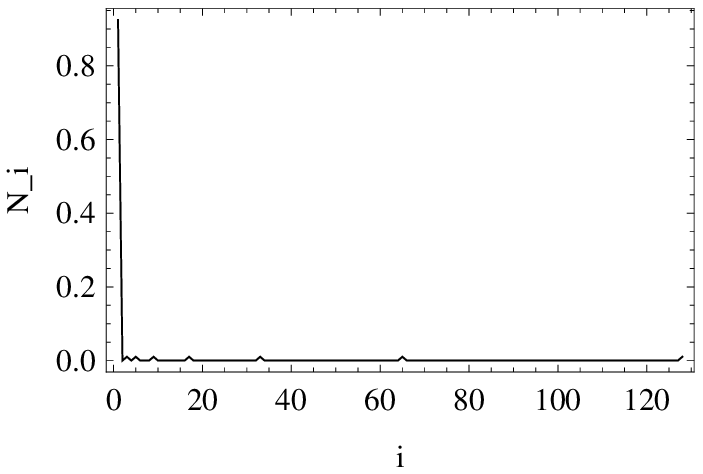}
\includegraphics[width=5.3cm]{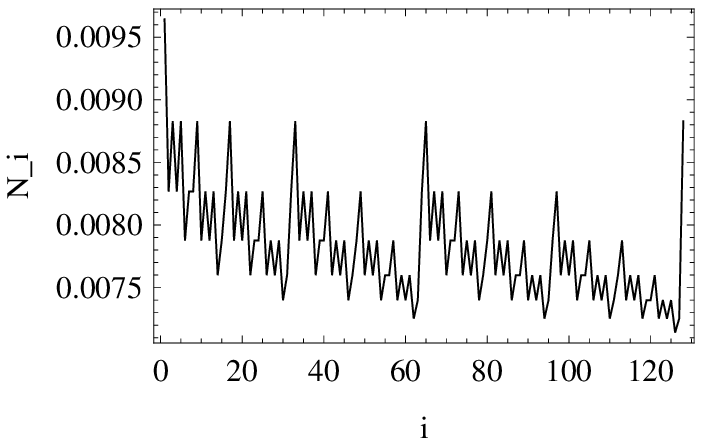}
\includegraphics[width=3.5cm]{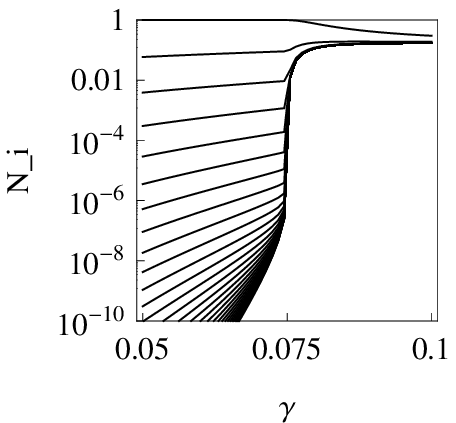}
\caption{\label{err_thr}Abundances $n_i^*$ of genotypes in the steady state in the single-peak model, for $L=7$ (which corresponds to $N=128$ genotypes), $\phi_1=10,\phi_2=\dots=\phi_N=1$, thus the fittest sequence being at $i=1$, and for two mutation rates $\gamma=0.01$ (left, well below the error threshold $\gamma_c(L=7)\approx 0.28$) and $\gamma=0.4$ (middle, above the threshold). Genotypes which correspond to the same Hamming distance from the master sequence $i=1$ (the same ``error class'') have the same abundance. Right: plot of $n_i$ for different error classes (lines from top to bottom) as a function of $\gamma$, for $L=30$. The transition, which is clearly visible at $\gamma_c(L=30) \approx 0.074$, becomes sharper for increasing $L$. }
\end{figure}

As already mentioned, the quasispecies is a set of closely related sequences which occupy a finite area in the genotype space. 
The name ``quasispecies'' comes from an apparent similarity to a real-world situation in which genotypes corresponding to organisms of the same species form a ``cloud'' of mutants in the genotype space around the best-fit genotype.
The quasispecies exists for any mutation rate $\gamma$ smaller than some critical $\gamma_c$, because in this case the steady state solution $\vec{n}^*\propto \vec{\psi}_1$ is localized around (usually) the maximal fitness, see Fig. \ref{err_thr}. Above $\gamma_c$, the steady state becomes delocalized and spreads over the whole genotype space. The critical $\gamma_c$ can be easily calculated \cite{nowak} for the single-peak fitness landscape:
\bq
	\gamma_c = 1 - \left(\frac{\phi_2}{\phi_1}\right)^{1/L}, \label{gammac}
\eq
where $\phi_1,\phi_2$ are fitnesses of the best-fit genotype and less-fit genotypes, respectively. Above $\gamma_c$, the best-fit genotype (the master sequence) vanishes from the population. This critical $\gamma_c$ is called the error threshold and the transition from localized to delocalized quasispecies is known as the error catastrophe. For increasing $L$ and $\phi_2/\phi_1$ kept constant, the critical mutation rate scales as $\gamma_c \sim 1/L$, thus longer sequences have smaller error thresholds.

\section{Quasispecies model with parallel mutation and selection}
\begin{figure}
	\centering
		\includegraphics[width=12cm]{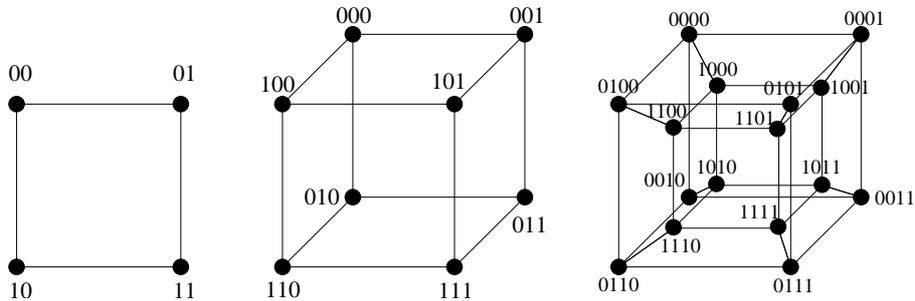}
	\caption{Examples of mutation graphs for the para-mu-se model, for $L=2,3,4$.}
	\label{fig:hypercube}
\end{figure}

The quasispecies model has an even simpler counterpart --- para-mu-se (parallel mutation and selection) model \cite{baake}. 
A key feature of this model in comparison to Eq.~(\ref{quasireact}) is that growth and mutation are decoupled:
\ba
	\mbox{genotype}\; i  &\xrightarrow{\phi_i - \gamma \sum_j A_{ij}}& 2\;\mbox{genotype}\; i,  \\
	\mbox{genotype}\; i  &\xrightarrow{\gamma A_{ij}}& \mbox{genotype}\; i + \mbox{genotype}\; j. 
\ea
Here $A$ is an adjacency matrix of a graph of possible mutations and $\gamma$ is the mutation rate. It is usually assumed that the matrix $A$ is symmetric (forward and reverse mutations have the same probability) and that $A_{ij} = 1$ if the Hamming distance $d(i,j)=1$. Therefore, mutations can change at most one symbol per replication. Then, $A$ is the adjacency matrix of the hypercube graph, see Fig. \ref{fig:hypercube}.

The model is known to have the error threshold for some fitness landscapes \cite{baake2}\footnote{For the single-peak landscape, eigenvector $\psi_1$ is always localized, but the net growth rate of the master sequence can be negative above some critical $\gamma_c$, which may be thought as a sort of the error catastrophe.} and for small $\gamma$ (for which $Q_{ij}\approx \gamma A_{ij}$ in Eq.~(\ref{qdef})) it behaves very much the same as the quasispecies model from Section 2. However, the para-mu-se version of the quasispecies model is generally simpler from a mathematical point of view. From now on, I will focus on this model, although general conclusions of this work remain valid also for the original quasispecies model.

In the limit of infinite populations, the para-mu-se model is described by the following set of equations (cf. Eq. (\ref{quasieq1})):
\bq
	\frac{{\rm d}}{{\rm d}t} n_i(t) = n_i(t) (\phi_i-J(t)) + \gamma \sum_{j=1}^N A_{ij} (n_j(t)-n_i(t)). \label{quasieq2}
\eq
Applying the same transformation (\ref{ntox}) as before, the above set is reduced to the linear form:
\bq
	\frac{{\rm d}}{{\rm d}t}\vec{x}(t) = W \vec{x}(t), \label{model}
\eq
where the matrix $W$ is now defined as $W_{ij} = \delta_{ij} \phi_i + \gamma\Delta_{ij}$, and $\Delta_{ij}=A_{ij}-\delta_{ij}\sum_k A_{ik }$ is the graph Laplacian. If $\lambda_1>\lambda_2>\dots>\lambda_N$ and $\{\vec{\psi_i}\}$ are again the eigenvalues and eigenvectors of $W$, the solution of Eq.~(\ref{model}) is given by the same Eq.~(\ref{quasiformal}) as for the quasispecies model, but with the new matrix $W$.

\section{Localization, random matrices and quasispecies}
The quasispecies or the para-mu-se model can be solved exactly in some special cases. The purpose of this article is to show that in the case of random, rugged fitness landscape which is much more difficult to study, there is an interesting analogy between the para-mu-se model, Anderson localization and RMT. Although the connection between localization theory and some evolutionary models in the continuous genotype space was made already more than 25 years ago \cite{engel}, the analogy (however obvious it may seem) between the para-mu-se model and RMT is not widely appreciated.

In what follows I will assume that the fitnesses $\{\phi_i\}$ are random numbers drawn from some distribution $p(\phi)$.
The quasispecies matrix $W$,
\bq
	W = \left[\begin{array}{cccc}
		\phi_1 & 0 & \dots & 0 \\
		0 & \phi_2 & \dots & 0 \\
		 & & \ddots & \\
		0 & \dots & 0 & \phi_N \\
	\end{array}\right] + \gamma \Delta = D +\gamma \Delta, \label{W_as_sum}
\eq
is the sum of a diagonal, random matrix $D={\rm diag}(\phi_1,\dots,\phi_N)$, and a Laplacian matrix $\gamma\Delta$.
The eigenproblem of $W$:
\bq
	\sum_j (\phi_{i}\delta_{ij}+\gamma \Delta_{ij})\psi_j = \lambda \psi_i
\eq
can be translated to the following Schr\"{o}dinger equation:
\bq
	-\sum_j \Delta_{ij}\psi_j + V_i \psi_i = E \psi_i,
\eq
where $V_i = -\phi_i/\gamma$ and $E=-\lambda/\gamma$. This is precisely the Anderson model of localization on arbitrary lattices (see, e.g., Ref. \cite{anderson2}) with a random potential $V_i$. High fitness values correspond to low potential values, and the ground state corresponds to the steady state $\vec{n}^* \propto \vec{\psi_1}$ (the quasispecies).

The matrix $W$ has random elements only on the diagonal, which makes it different from what is usually considered to be a random matrix --- a matrix with a finite fraction of elements being (possibly correlated) random numbers. Such ``dense'' random matrices, which form the core of random matrix theory, appear in many problems in physics \cite{guhr}, telecommunication and information theory \cite{moustakas}, and quantitative finance \cite{laloux}. However, matrices in which only diagonal elements (or elements in a narrow band) are random, are also quite abundant in physics, in particular in quantum chaos \cite{izrailev, mirlin, zyczkowski} and Anderson localization problems \cite{biroli, metz, gudowska, neu, biroli2}. The matrix $W$ with random fitness values defined above belongs to the same class of sparse random matrices. There is, however, one important difference: typical systems studied in the framework of localization theory are usually low-dimensional, except for Bethe lattices \cite{anderson2, biroli}. The reason is that low-dimensional systems can model real physical situations like transport properties of disordered solids \cite{kramer}. In addition, many analytical results have been obtained for 1d or Bethe lattices due to their special, simple structure. In contrast, the quasispecies problem is multi-dimensional, since the Laplacian $\Delta$ is defined on the hypercube graph, and the matrix $W$, albeit sparse, has a non-trivial structure. Although this can generally make analytical calculations very hard, it will not be relevant to the problems studied in this work.

Random matrix theory deals primarily with eigenvalues and eigenvectors of matrices. Table 1 lists some of typical quantities calculated in RMT. It turns out that some of them are directly related to quantities relevant to the quasispecies theory. The first example I shall consider are spectral properties of the matrix $W$. A simple reasoning shows that differences between nearest eigenvalues of $W$ determine typical timescales in the model. In the limit of small mutation rate $\gamma\approx 0$, $W$ is almost diagonal and the eigenvalues $\{\lambda_i\}$ are equal to the fitnesses $\{\phi_i\}$. For simplicity, let the fitnesses decrease with $i$, so that the ordered eigenvalues are $\lambda_i\cong \phi_i$. All eigenvectors are then trivially localized: $\psi_{i,j}\cong \delta_{ij}$, and the best adapted genotype corresponds to the eigenvector $\vec{\psi_1}$, the second best adapted one to the eigenvector  $\vec{\psi_2}$, and so on. If the population is initially localized at the least adapted genotype $N$, then scalar products $\vec{\psi_i}\cdot\vec{x}(0)$ decay exponentially fast with increasing Hamming distance $d(N,i)$. Writing the solution of Eq.~(\ref{model}) as
\ba
	\vec{x}(t) &=& e^{tW}\vec{x}(0)= \sum_i e^{t\lambda_i} \left(\vec{\psi_i}\vec{x}(0)\right) \vec{\psi}_i \nonumber \\
	&=& e^{t\lambda_1} \left[ \left(\vec{\psi_1}\vec{x}(0)\right) \vec{\psi}_1 + e^{-t(\lambda_1-\lambda_2)}\left[ \left(\vec{\psi_2}\vec{x}(0)\right) \vec{\psi}_2 + e^{-t(\lambda_2-\lambda_3)}\left[ \left(\vec{\psi_3}\vec{x}(0)\right) \vec{\psi}_3+\dots\right]\right]\right] \nonumber
\ea
reveals that the contribution of the eigenvectors $\vec{\psi}_N,\vec{\psi}_{N-1},\dots,\vec{\psi_2}$ to $\vec{x}(t)$ first increase (in this order) and then decay with rates $\lambda_{N-1}-\lambda_{N},\dots,\lambda_2-\lambda_3,\lambda_1-\lambda_2$. The rates $\lambda_{i}-\lambda_{i+1}$ gives different timescales in the system. In particular, 
\bq
	\tau = \frac{1}{\lambda_1-\lambda_2} 
\eq
is the characteristic time to reach the steady state $\vec{n}^* \propto \vec{\psi}_1$. Other differences $\lambda_{i}-\lambda_{i+1}$ correspond to characteristic times
\bq
	\tau_i = \frac{1}{\lambda_{i}-\lambda_{i+1}} 
\eq 
related to ``jumps'' between locally adapted quasispecies (Fig. \ref{fig:wigner_surmise}, left). If the distribution of differences $\lambda_{i}-\lambda_{i+1}$ is known, one can calculate the average time between these events, and hence estimate the speed of evolution. However, the probability distribution of differences $S_n(s)$ with $s=\lambda_n-\lambda_{n+1}$ is just the nearest-neighbour spacing distribution, which is very commonly used in RMT to study short-range fluctuations in the spectrum (Fig. \ref{fig:wigner_surmise}, right). Therefore, the problem of time evolution in the quasispecies theory is equivalent to the problem of finding $S_n(s)$ for a particular ensemble of random matrices $W$. The rest of the paper is devoted to calculating this quantity and interpreting it from a quasispecies perspective.

\begin{table}[h]
	\caption{\label{tab1}Correspondence between quantities of interest in RMT and in the quasispecies theory}
	\vspace{5mm}
	\centering
		\begin{tabular}{ccc}
			\hline
			{\bf RMT} & & {\bf Quasispecies} \\			
			level spacing & $\leftrightarrow$ & statistics of jumps in fitness space \\
			distribution of maximal eigenvalue & $\leftrightarrow$ & steady-state total abundance \\
			localization of eigenvectors & $\leftrightarrow$ & error threshold \\
			participation ratio of eigenvectors & $\leftrightarrow$ & genetic diversity \\
			\hline
		\end{tabular}
\end{table}

Before I proceed to calculations of $S_n(s)$, I will briefly mention other similarities between the quasispecies theory and RMT. For example, the formula for the steady-state abundances $\vec{n}^*$ from Eq.~(\ref{vecnn}) shows that the maximal eigenvalue $\lambda_1$ plays the role of the total abundance of all possible genotypes. The distribution of the maximal eigenvalue is frequently studied in the framework of RMT.

Finally, the statistics of eigenvectors $\{\vec{\psi}_i\}$ of some random matrices turns out to be very important for localization problems. In particular, participation ratio of eigenvectors is studied as a measure of localization. However, the localization transition in matrix models corresponds to the error catastrophe in the quasispecies model, which shows yet another interesting connection between RMT and the quasispecies theory. Table \ref{tab1} provides a summary of all analogies mentioned in this work.

\section{Level spacing distribution}
In this section I will discuss the level spacing distribution $S_n(s)$, and its average over all nearest-neighbour pairs of eigenvalues 
\bq
	S(s)\equiv \frac{1}{N-1}\sum_{n=1}^{N-1} S_n(s)
\eq
for the matrix $W$ in the para-mu-se model. I will assume the uniform distribution of fitness: $p(\phi)=1$ in the range $0\dots 1$. This ensures that there is always a maximal and a minimal fitness, which is biologically relevant (the growth rate cannot be arbitrarily large), and that for large $N=2^L$, ${\rm max}\{\phi_1,\dots,\phi_N\} \to 1$ is bounded from above\footnote{Note that for $p(\phi) = e^{-\phi}$ which is often assumed by various authors \cite{krug-karl,jain}, ${\rm max}\{\phi_i\} \sim L$.}. At last, the uniform distribution of $\phi_i$ allows one to draw yet another link to Anderson localization, in which site potentials are also uniformly distributed \cite{anderson}.

\begin{figure}
\psfrag{Ps}{$S(s)$} \psfrag{s}{$s$}
\psfrag{ts}{$t=1/s$}
	\centering
		\includegraphics[width=6cm]{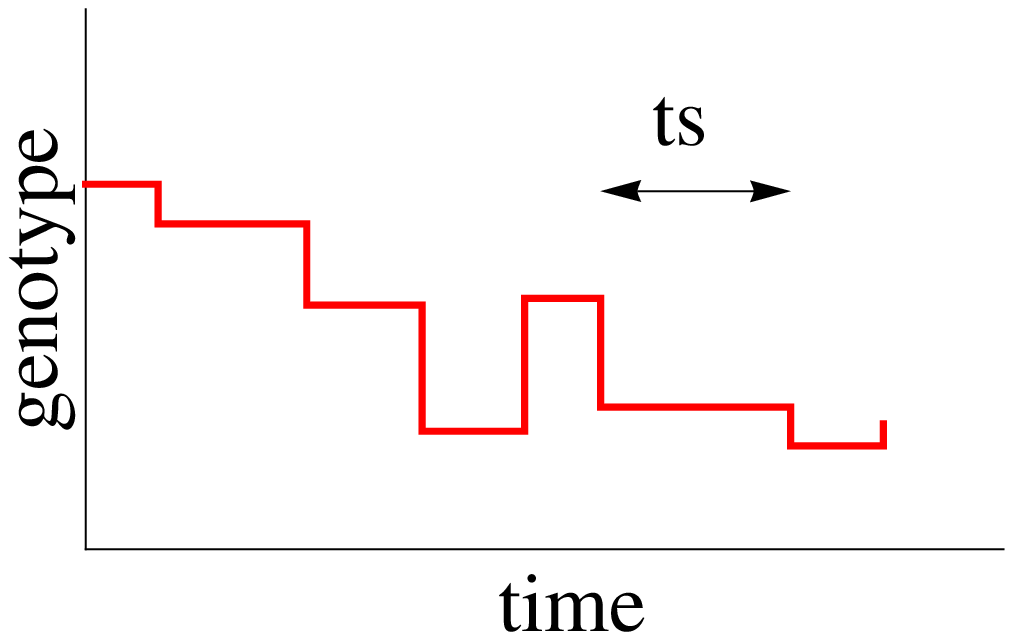}
		\includegraphics[width=6cm]{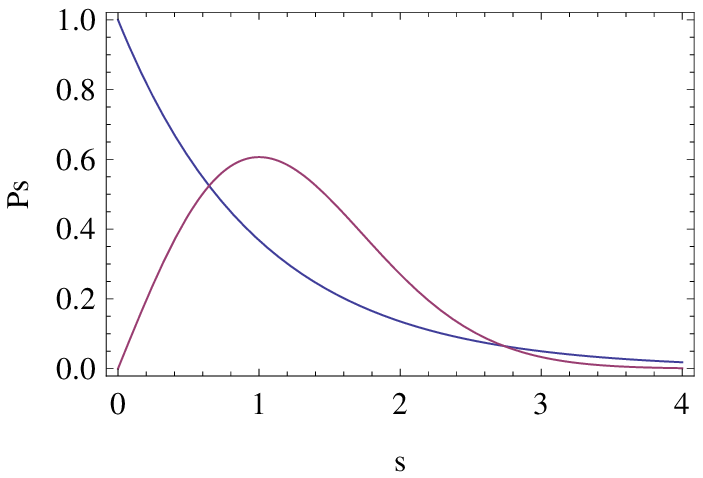}
	\caption{Left: a schematic view of jumps in the fitness space. Red line shows the trajectory of the most populated genotype. Typical timescale $t$ is inversely proportional to the separation $s$ of eigenvalues of $W$. Right: two level spacing distributions -- exponential and Wigner's surmise -- frequently encountered in RMT.}
	\label{fig:wigner_surmise}
\end{figure}

As explained above, dynamical properties of the quasispecies or the para-mu-se model can be inferred from the distribution $S_n(s)$. 
One of the most important results of RMT is that eigenvalues usually ``repel'' each other in the spectrum \cite{guhr} so that $S(s)$ is zero at $s=0$. In particular, for Gaussian random matrices, we have to a good approximation
\bq
	S(s) \approx s e^{-s^2/2},
\eq
which is known as the Wigner surmise. This level repulsion is characteristic for interacting systems, in which eigenvalues are correlated. For uncorrelated eigenvalues, the level-spacing distribution is exponential, $S(s) = e^{-s}$, for the unfolded spectrum, i.e., after transforming all eigenvalues such that their spectral density is uniform. Both the Wigner surmise and the exponential distribution are plotted in Fig. \ref{fig:wigner_surmise}, right.

\begin{figure}
	\centering
	\psfrag{Ps}{$S(s)$} \psfrag{s}{$s$}
		\includegraphics[width=6cm]{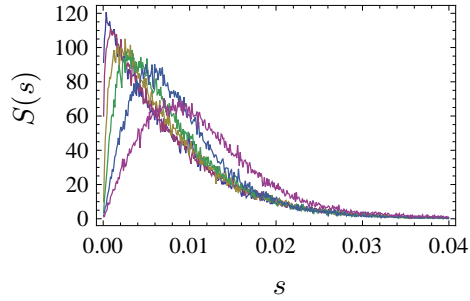}
	\caption{Level-spacing distribution $S(s)$ for $L=7$ and $\gamma=0.005,0.01,0.02,0.03,0.05,0.1$ (curves from left to right).}
	\label{fig:Ps1}
\end{figure}

\begin{figure}
	\psfrag{s}{\footnotesize $Ns$}
	\psfrag{Ps}{\footnotesize \hspace{-5mm}$S(s)/N$}
	\centering
		\includegraphics[width=4.8cm]{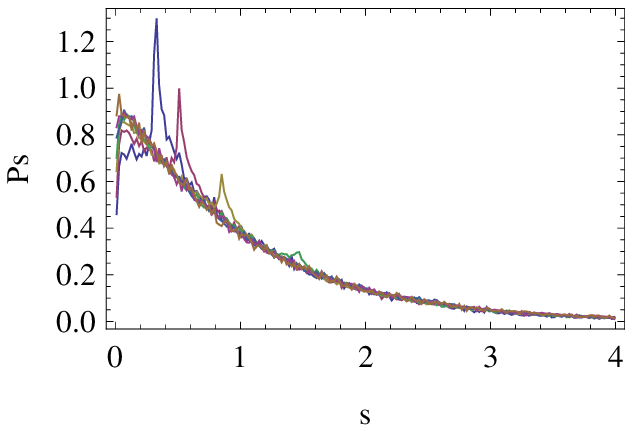}
		\includegraphics[width=4.8cm]{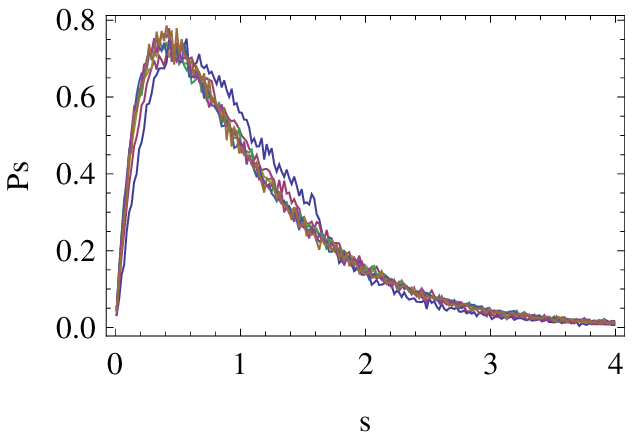}
		\includegraphics[width=4.8cm]{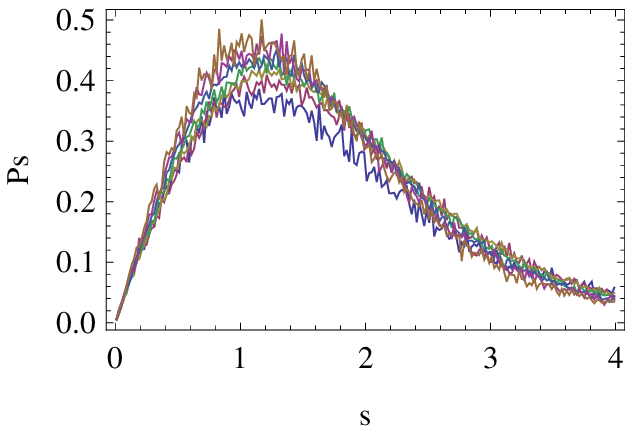}
	\caption{Rescaled $S(s)$ for $L=4,5,6,7,8,9,10$, and $\gamma=g/L$, where $g=0.04$ (left), $0.2$ (middle) and $1$ (right). For $g=0.04$ some spikes are visible for $s=2\gamma N$. These spikes are finite-size effects, which vanish with increasing $L$.}
	\label{fig:Presc1}
\end{figure}

It is evident from Eq.~(\ref{W_as_sum}) that the mutation rate $\gamma$ plays the role of interaction strength, so we expect that $S(s)$ should be exponential for $\gamma\to 0$, and that it will show signs of level repulsion for $\gamma>0$. This is indeed seen in Fig. \ref{fig:Ps1}, in which I plot $S(s)$ obtained from numerical diagonalization of $W$ for uniform fitness distribution and various mutation rates.
A simple argument shows that for $L$ large enough,
\bq
	S(s) \cong Nf(\gamma L,Ns), \label{Ps_scaling}
\eq
where $f(g,x)$ is a (yet unknown) semi-positive function. This means that $S(s)$ obtained for different lengths $L$ and for mutation rates $\gamma = g/L$ with some arbitrary $g$ should collapse to a single curve when plotted in the rescaled variable $x=Ns$, i.e., when we ``blow up'' the spectrum of eigenvalues.
This can indeed be seen in Fig. \ref{fig:Presc1}. The scaling form (\ref{Ps_scaling}) has a simple motivation. Firstly, the number of eigenvalues of the matrix $W$ is $N$. Since $W=D+\gamma \Delta$, for small $\gamma$ we expect that the spectrum of $W$ will have a similar width as the spectrum of $D$. But the spectrum $\rho_D(\lambda)$ of the diagonal matrix $D$ reads $\rho_D(\lambda) = p(\lambda)$, where $p(\phi)=1$ for $0\le \phi\le 1$, and it has a finite support of length one. Therefore, the average distance between the eigenvalues of $W$ must scale as $1/N$, which explains the factor $N$ blowing up the nearest-level spacing in Eq. (\ref{Ps_scaling}).

Secondly, the net growth rate of genotype $i$ is $\phi_i-L\gamma +\gamma \sum_j A_{ij} n_j/n_i = \phi_i + O(\gamma L)$, thus $\gamma$ appears in the para-mu-se equations (\ref{quasieq2}) always as a product of $L$ and $\gamma$. We thus expect that $\gamma L$ should be the relevant variable for the balance between growth and mutation. To make it more explicit, we observe that for small $\gamma$, the quasispecies is localized at the largest fitness, which for the uniform distribution $p(\phi)$ is $\phi_1 \approx 1$. This best adapted genotype is surrounded by a sea of less adapted genotypes with average fitnesses $\phi_2\approx 1/2$. The relative abundances of the best adapted genotype, $x_1$, and a less-adapted one, say, $x_2$, can be approximately determined from
\ba
	\dot{x}_1(t) &=& x_1(t) (\phi_1 - L\gamma) + L\gamma x_2(t), \\
	\dot{x}_2(t) &=& x_2(t) \phi_2 + \gamma x_1(t).
\ea
Here I have assumed that mutations of the less-adapted genotype produce mainly genotypes from the same, less-adapted class of genotypes, therefore there is no loss term due to mutations in the second equation. Then, the difference of eigenvalues of the corresponding $2\times 2$ matrix $W$ reads
\bq
	\lambda_1-\lambda_2 = \sqrt{L^2 \gamma^2 + (\phi_1 - \phi_2)^2 + 
 2 L \gamma (2 \gamma - \phi_1 + \phi_2)},
\eq
and it is evident that (assuming that $2\gamma\ll \phi_1-\phi_2 \approx 1/2$) $s=\lambda_1-\lambda_2$ depends only on the product of $L\gamma$ in the limit of small mutation rate.

The $N$-scaling from Eq.~(\ref{Ps_scaling}) can be deduced analytically for uniform $p(\phi)$ in the limit of $\gamma\to 0$. In this limit, as already mentioned, the eigenvalues of $W$ are distributed uniformly between 0 and 1. This means that, effectively, there are no interactions in the system, so $S(s)$ for the unfolded spectrum should be exponential. This is very easy to check. The probability $S_n(s)$ that the difference for an ordered pair of eigenvalues $\lambda_n>\lambda_{n+1}$ will be $s$ is given by
\bq
	S_n(s) = Z^{-1} \int_{-\infty}^\infty d\lambda_1 \int_{\lambda_1}^\infty d\lambda_2 \cdots \int_{\lambda_{N-1}}^\infty d\lambda_N p(\lambda_1)\cdots p(\lambda_N) \delta(\lambda_{n+1}-\lambda_n - s), \label{Ps}
\eq
where
\bq
	Z= \int_{-\infty}^\infty d\lambda_1 \int_{\lambda_1}^\infty d\lambda_2 \cdots \int_{\lambda_{N-1}}^\infty d\lambda_N p(\lambda_1)\cdots p(\lambda_N).
	\label{ZN}
\eq
Equations (\ref{Ps}) and (\ref{ZN}) can be evaluated for the uniform fitness distribution. We obtain $Z=1/N!$ and
\bq
	S_n(s) = N(1-s)^N \approx Ne^{-Ns}, \label{Ps_uni}
\eq
which does not depend on $n$, thus the average level-spacing is also $S(s)\cong Ne^{-Ns}$. For large $N$, $S(s)$ scales as in Eq.~(\ref{Ps_scaling}) with $\gamma=0$ and
\bq
	f(x)=e^{-x}. \label{f_uni}
\eq
Such an exponential decay is indeed visible in Fig. \ref{fig:Presc1}, left.

\begin{figure}
	\psfrag{s}{\footnotesize $Ns$}
	\psfrag{Pmaxs}{\footnotesize $S_1(s)/N$}

	\centering
		\includegraphics[width=4.8cm]{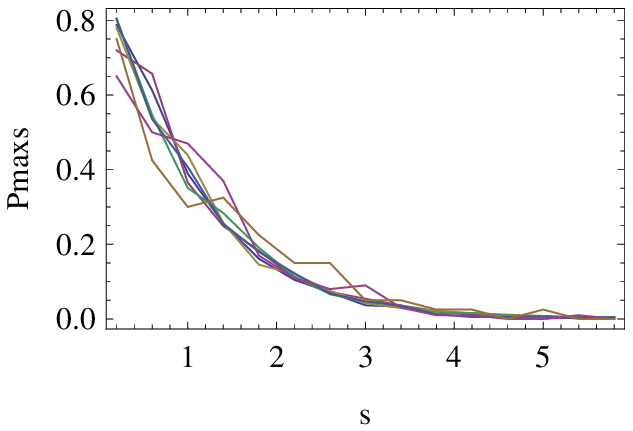}
		\includegraphics[width=4.8cm]{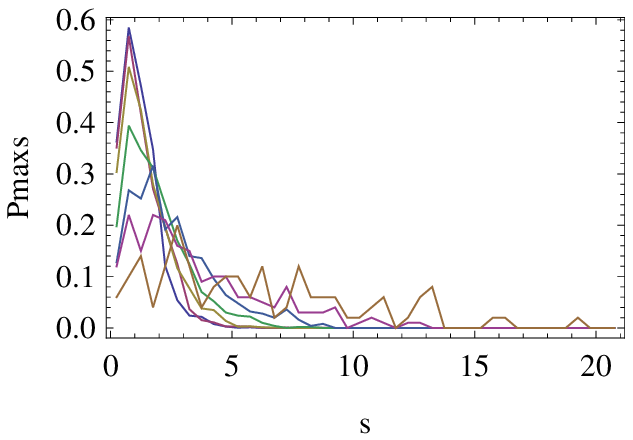}
		\includegraphics[width=4.8cm]{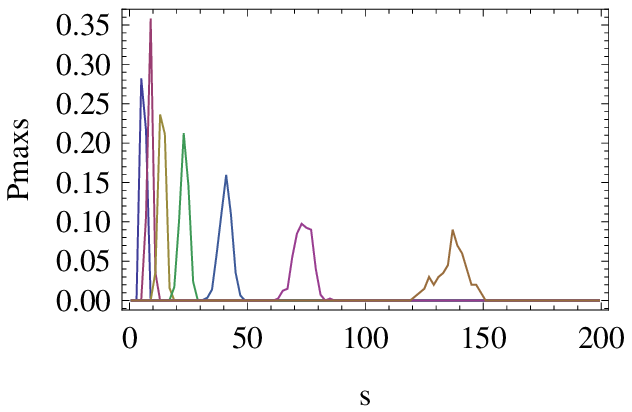}
	\caption{Rescaled distribution of the gap $S_{1}(s)$ between two largest eigenvalues, for $L=4,\dots,10$, $\gamma=g/L$, and $g=0.04$ (left), $g=0.2$ (middle) and $g=1.0$ (right). The distribution shifts to the right for increasing $L$. }
	\label{fig:pmax}
\end{figure}

\begin{figure}
	\psfrag{PR}{\footnotesize $PR/2^L$}\psfrag{L}{\footnotesize $L$}
	\centering
		\includegraphics[width=6cm]{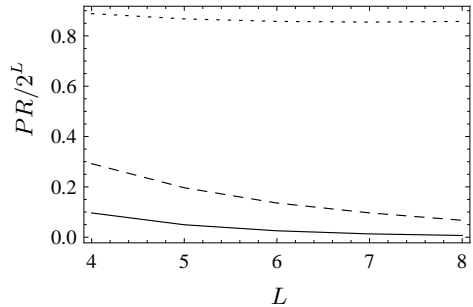}
	\caption{Participation ratio divided by $N=2^L$ for $g=0.04$ (solid), $g=0.2$ (dashed) and $g=1.0$ (dotted).}
	\label{fig:pr}
\end{figure}

Numerical simulations presented in Fig. \ref{fig:Presc1} indicate that the scaling form (\ref{Ps_scaling}) is valid also for $\gamma\equiv g/L>0$, i.e., when $\gamma$ scales inversely with the length $L$ of the sequence. For $g$ small enough, the same scaling holds for the distribution $S_{1}(s)$ of the gap between two largest eigenvalues $s=\lambda_1-\lambda_2$, see Fig. \ref{fig:pmax}, left. However, for large $g$, the distributions shift to larger $s$ with increasing $L$ (Fig. \ref{fig:pmax} middle and right). Figure \ref{fig:pr} shows plots of the participation ratio divided by the total number of genotypes, 
\bq
	PR = \frac{1}{2^L} \left(\sum_i \psi_{1,i}\right)^2 / \sum_i \psi_{1,i}^2,
\eq
which measures how strongly the quasispecies is localized: $PR\approx 0$ for eigenvectors with only few entries larger than zero, whereas $PR\approx 1$ means that all entries are roughly the same. In Fig.~\ref{fig:pr}, $PR\approx 0$ for small $g$, but for sufficiently large $g$, $PR$ is of order one. This means that the principal eigenvector covers a finite fraction of the genotype space, hence the quasispecies is no longer localized. This indicates a transition similar to the error catastrophe in the quasispecies model.

\section{Statistics of jumps}
Using the scaling hypothesis for the level-spacing distribution which, as we have seen, is valid for any mutation rate $\gamma$ small enough (so that the quasispecies remains localized), the statistics of jumps can be inferred very easily. The distribution $P_{\rm jumps}(\tau)$ of times between  jumps is given by
\bq
	P_{\rm jumps}(\tau) = \int_0^\infty S(s) \delta(\tau-1/s) ds = N\frac{1}{\tau^2}f(\gamma L, N/\tau). \label{ptau}
\eq
The same formula holds for $P_{\rm ss}(\tau)$, the distribution of times to steady state, because the gap distribution $S_1(s) \approx S(s)$ for localized quasispecies. This means that the mean time to reach the steady state, which is the time it takes until the quasispecies stops evolving, is
\bq
	\left<\tau\right> = N \int_0^\infty \frac{1}{x} f(\gamma L,x) dx,
\eq
and it grows linearly with $N$ (or exponentially with the length of the sequence $L$). For $\gamma=0$, the above integral is divergent, because $f(\gamma L,0)>0$. This is correct, because in the absence of mutations there is no way to reach the steady state from any other point in the genotype space. But for $\gamma>0$, we have $f(\gamma L,0)=0$ due to level repulsion. For small $x$, $f(\gamma L,x)$ is proportional to $x$ as it can be seen in Fig. \ref{fig:Presc1}. Therefore, $\int_0^\infty \frac{1}{x} f(\gamma L,x) dx<\infty$, and the average time $\left<\tau\right>$ is finite. 

\begin{figure}
	\psfrag{tau}{\footnotesize $\tau$}\psfrag{C}{\footnotesize $C(\tau)$}
	\includegraphics[width=7cm]{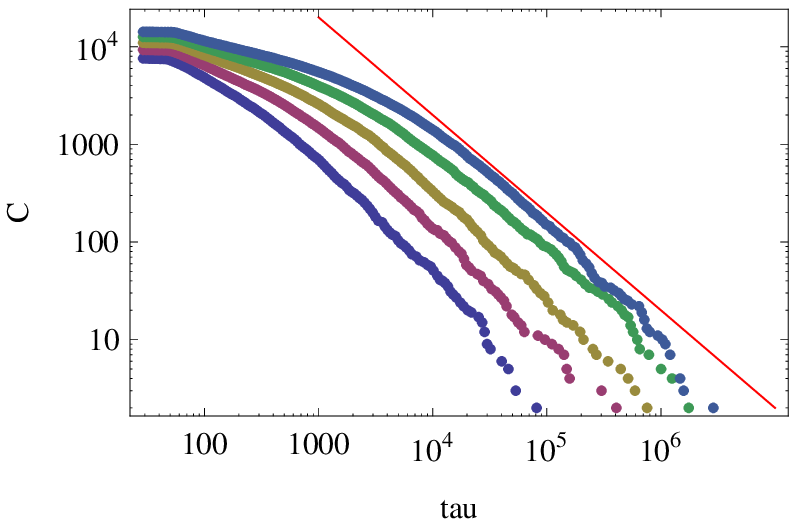}
	\includegraphics[width=7cm]{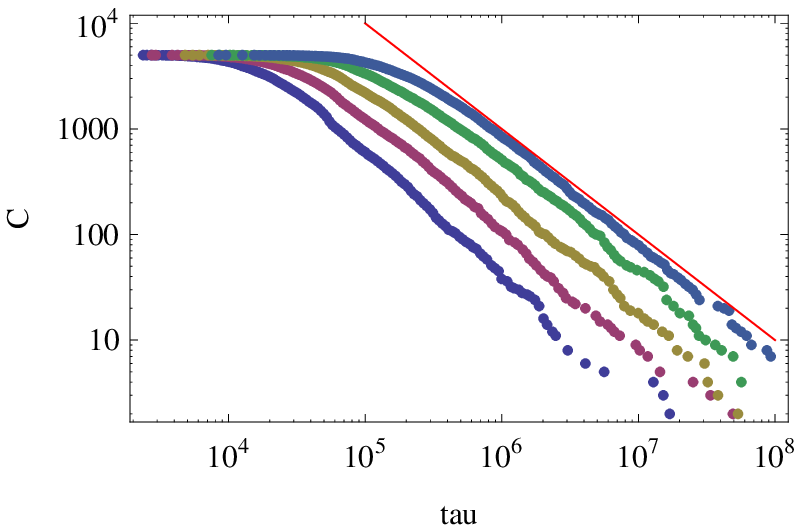}
	\caption{\label{fig:jumps}Left: Plots of the cumulative distribution of jumps $C_{\rm jumps}(\tau) = \int_{\tau}^\infty P_{\rm jumps}(\tau') d\tau'$ for $L=4,5,6,7,8$ (from left to right) and $\gamma=0.04/L$. Solid line corresponds to the theoretical result $C(\tau)\sim \tau^{-1}$. Right: $C_{\rm ss}(\tau)$ for the time to reach steady state. }
\end{figure}

Equation (\ref{ptau}) tells us that for $N\ll \tau \ll N/\gamma$, for which $f(\gamma L,x)\approx \rm{const}$, the distribution of times between jumps follows a power law: $P_{\rm jumps}(\tau)\sim \tau^{-2}$. This is also true for the time to reach the steady state, $P_{\rm ss}(\tau)\sim \tau^{-2}$. In Fig. \ref{fig:jumps} I show plots of the cumulative distribution of times between jumps measured directly by solving differential equations (\ref{quasieq2}) numerically for 5000 random fitness landscapes, and tracing the position of the maximal abundance $n_i$. A jump was recorded whenever this maximal abundance changed its location in the genotype space. In this way, the statistics of times between jumps was obtained. For each fitness landscape, the simulation was stopped when the difference $\sum_i |n_i^* - n_i(t)|/N$ between the steady-state solution and $n_i(t)$ was smaller than $10^{-6}$. In this way, also the statistics of times to steady state was collected. The experimental cumulative distributions $C_{\rm jumps}(\tau) = \int_{\tau}^\infty P_{\rm jumps}(\tau') d\tau'$ and $C_{\rm ss}(\tau) = \int_{\tau}^\infty P_{\rm ss}(\tau') d\tau'$ presented in Fig. \ref{fig:jumps} show a power-law behaviour with an exponent close to minus one, $C_{\rm jumps}(\tau)\sim C_{\rm ss}(\tau)\sim \tau^{-1}$, in good agreement with theory.

The power-law behaviour of $P_{\rm jumps}(\tau)$ for individual jumps as well $P_{\rm ss}(\tau)$ for the time to reach steady state has been observed in a simplified ``shell'' model in the strong selection limit \cite{krug-karl,jain2}, which would correspond to $\gamma\to 0$ in our model. We see that a simple observation, which relates the dynamics of the quasispecies to the level spacing distribution, allows one to extend this result to the case of $\gamma=g/L >0$.

\section{Conclusion}
The main objective of this work was to present an interesting analogy between the quasispecies theory (in particular, the para-mu-se model) and random matrices which appear in localization theory. 
I discussed how static and dynamical properties of the quasispecies model are related to quantities such as nearest-level spacing distribution or participation ratio of eigenvectors, which are typically calculated within the framework of random matrix theory. Although most of the results presented here were obtained in numerical simulations, they were all corroborated by simple, mathematical calculations. It remains a challenge to calculate analytically the level-spacing distribution for non-zero mutation rates.

The analogy to Anderson localization mentioned here has been already mentioned in Refs.~\cite{epstein,engel} and, more recently, in Ref.~\cite{BW}, in which some results of localization theory for 1d tight-binding models are used to find the point of the phase transition in a model with two quasispecies linked by migration. However, no systematic studies of localization on the hypercube with random distribution of fitness (site potential) have been made so far. This would be potentially a very interesting research area which could further link biological evolution models, localization theory, and random matrices.

\section*{Acknowledgments}
I thank Z. Burda, M.R. Evans, and J.-M. Luck for many valuable discussions. I am particularly indebted to R. Allen for introducing me to problems of biological (especially microbial) evolution. I also acknowledge support by the EPSRC under grant EP/E030173.

\end{document}